\index{\label{} }
\documentclass[manuscript]{aastex}
\usepackage{textcomp}
\usepackage{lscape}
\usepackage{multicol}
\usepackage{epsfig}

\slugcomment{Not to appear in Nonlearned J., 45.}

\shorttitle{Optical Imaging and Spectroscopic Observation of the
Galactic Supernova Remnant G85.9-0.6}

\begin{document}


\title{Optical Imaging and Spectroscopic Observation of the
Galactic Supernova Remnant G85.9-0.6}

\author{F.G\"{o}k\altaffilmark{1} , A.Sezer\altaffilmark{2, 4} ,
E.Aktekin\altaffilmark{1}, T.G\"{u}ver\altaffilmark{3} and
N.Ercan\altaffilmark{4}}

\affil{Akdeniz University, Faculty of Art and Sciences, Department
of Physics\\ Antalya, 07058, Turkey}

\affil{T\"UB\.ITAK Space Technologies Research Institute,
Middle-East Technical University, 06531 Ankara, Turkey}

\affil{University of Arizona, Department of Astronomy, 933 N.
Cherry Ave., Tucson, AZ, 85721}

\affil{Bo\~gazi\d{c}i University, Faculty of Art and Sciences,
Department of Physics\\
\.Istanbul, 34342, Turkey}

\begin{abstract}

Optical CCD imaging with H$\alpha$ and [SII] filters and
spectroscopic observations of the galactic supernova remnant
G85.9-0.6 have been performed for the first time. The CCD image
data are taken with the 1.5m Russian-Turkish Telescope (RTT150) at
T\"{U}B\.{I}TAK National Observatory (TUG) and spectral data are
taken with the Bok 2.3 m telescope on Kitt Peak, AZ.

The images are taken with narrow-band interference filters
H$\alpha$, [SII] and their continuum. [SII]/H$\alpha$ ratio image
is performed. The ratio obtained from [SII]/H$\alpha$ is found to
be $\sim$0.42, indicating that the remnant interacts with HII
regions. G85.9-0.6 shows diffuse-shell morphology.
[SII]$\lambda\lambda 6716/6731$ average flux ratio is calculated
from the spectra, and the electron density $N_{e}$ is obtained to
be 395 $cm^{-3}$. From [OIII]/H$\beta$ ratio, shock velocity has
been estimated, pre-shock density of $n_{c}=14$ $cm^{-3}$,
explosion energy of $E=9.2\times10^{50}$ ergs, interstellar
extinction of $E(B-V)=0.28$, and neutral hydrogen column density
of $N(HI)=1.53\times10^{21}$ $cm^{-2}$ are reported.

\end{abstract}

\keywords{G85.9-0.6; optical observation; CCD image; spectra}

\section{Introduction}

Galactic supernova remnants (SNRs) are now as many as 274
\citep{gre09}. Most of them are observed in radio wavelengths from
their non-thermal synchrotron emission despite their very few
optical observations. The reason for this is because of the dust
and gas around the galactic disk stopping the optical light coming
from them. However, only recently optical observations of SNRs can
be performed by using narrow-band interference filters and
providing rather long posing intervals \citep[e.g.][]{fes97,
fes08,mav01,mav03,bou02,bou09,gok08}. Recently, X-ray satellites
like Chandra, XMM-Newton, Suzaku observed some SNRs in X-ray band
\citep [e.g.][] {bam00,kat09}.

The galactic supernova remnant G85.9-0.6 ($\alpha=20^{h} 58^{m}
40^{s}$, $\delta= 44^{0} 53^{'}$) was discovered in the Canadian
Galactic Plane Survey (CGPS) \citep{kot01}. G85.9-0.6 has a radio
surface brightness at 1 GHz of $2\times 10^{-22}$ $Watt m^{-2}
Hz^{-1} sr^{-1}$. \citet{kot01} has detected no HI features
corresponding to the SNR. This and the low radio and X-ray
brightness, suggest expansion in a low-density medium. The SNR may
lie in the low-density region between the local and Perseus spiral
arms, at a distance of about 5 kpc. The most likely event that
would produce an SNR in such a low density medium might be a Type
Ia supernova. It has a radius of $\sim 12^{'}$ \citep{kot01}.
G85.9-0.6 has a diffuse type structure in X-ray region, where its
electron density is $0.20$ $cm^{-3}$ and the ejecta mass is at
order of $1.0$ $M_{\odot}$ \citep{jac08}.

[SII]/H$\alpha$ ratio is an important criteria to distinguish HII
regions and SNRs \citep{mat72, dop97}. In HII regions most sulphur
atoms are in the form of SIII, since there is a strong
photo-ionization and [SII]/H$\alpha$ ratio is expected to be in
the range of $\sim 0.1-0.3$ \citep{ost89}, however recent models
give this range to be 0.1-0.5 \citep{dop00, kew01}, for SNRs on
the other hand, it is 0.5-1.0 \citep{ray79, shu79}.

In this work, we selected the SNRs (G16.2-2.7, G17.8-2.6,
G36.6+2.6, G83.0-0.3 and G85.9-0.6) that have no optical
observation in literature so far and have proper coordinates to
observe at TUG. We take their H$\alpha$ images with 300 seconds
exposure time.

Theoretical models predict that, in SNRs, the temperature of the
regions that give optical emission should be at $\sim10^{4}$ $K$,
therefore, the intensity of H$\alpha$ emission line should be
larger compared to the other emission lines. For this reason,
H$\alpha$ interference filters are preferred for determining SNRs
\citep{ost89}. With H$\alpha$ filter we found radiation only from
G85.9-0.6 and studied its optical image and spectrum. The
observations and results are presented in sections 2. and 3.,
respectively. In the last part we discuss the results.

\section{Observations}
\subsection[]{Imaging}
The CCD images of G85.9-0.6 are obtained with the 1.5m RTT150 at
TUG, Antalya, T\"{u}rkiye. The data are taken with a
2048$\times$2048 pixel CCD447, with a pixel size of 15$\mu m$
resulting in a 13.5 arcmin $\times$ 13.5 arcmin field of view,
with a plate scale of $0^{''}.39$ $pixel^{-1}$. Images are
obtained with H$\alpha$, [SII] and their continuum filters within
900 seconds exposure times. Continuum filters are used to
eliminate contamination from background starlight. Bias frames and
dome flats are also observed for data reduction. The observations
and interference filter characteristics are given in Table 1. The
coordinates quoted in this work refer to epoch 2000. The image
reduction is performed by using standard IRAF (Image Reduction
Analysis Facility) routines.

Backgrounded, flat-fielded, trimmed and continuum subtracted [SII]
and H$\alpha$ images are prepared. In Figure 1, H$\alpha$,
H$\alpha$ cont., H$\alpha $-H$\alpha $ cont. and [SII], [SII]
cont., [SII]-[SII] cont. images are presented. The images have
been smoothed to suppress the remaining of the imperfect continuum
subtraction.

\subsection[]{Spectroscopy}
An optical spectrum of SNR G85.9-0.6, covering 3750 - 7250 $\AA$
range, was obtained on UT 2009 June 16 at the Bok 2.3 m telescope
on Kitt Peak, AZ, using the Boller and Chivens Spectrograph at the
Richey-Chretien f/ 9 focus together with a thinned
back-illuminated 1200$\times$800, 15$\mu m$ pixel Loral CCD. We
used a 1st order 400 g/mm grating blazed at 4889 $\AA$. With a
slit width of 2.5 arcseconds a typical resolution of 6 $\AA$ was
achieved at a reciprocal dispersion of 2.78 $\AA$ / pix on the
CCD. Spectrums are taken on the image from three different
intensive emission regions far from the stars. The coordinates of
these regions and exposure times are listed in Table 2.

The data reduction is carried out by using the standard IRAF
routines. Each frame is bias subtracted and for flat- field
correction halogen lamps are used. Then with {\it apall} package
under {\it noao.twodspec.apextract}, each spectrum is cleaned from
external factors like sky and cosmic rays. The {\it apall} package
selects an aperture for the target spectrum, and two apertures for
the sky spectrum. The apertures for the sky spectrum are selected
a dozen or so pixels, not illuminated by the object, above and
below the target spectrum. {\it Apall} takes the average of the
two sky spectrum and subtracts it from the target spectrum giving
a sky subtracted target spectrum. {\it Identify}, {\it
reidentify}, {\it hedit} and {\it dispcor} packages under {\it
noao.onedspec} are used for wavelength calibration. For flux
calibration, {\it sensfunc} package under {\it noao.onedspec} and
{\it calibrate} package under {\it noao.twodspec.longslit} are
used as described by \citet{mas97}. The spectra is dereddened by
using {\it dereddened} task under {\it noao.onedspec}. The
spectrophotometric standard star HR5501 is observed for flux
calibration of the spectra \citep{ham92}. For wavelength
calibration, however, helium-neon (He-Ne) arc lamps are used.

The spectra obtained for three region are shown in Figure 2. In
the optical spectra of SNRs; H$\alpha$, H$\beta$ and
[OIII]$\lambda 4959,\lambda 5007$, [NII]$\lambda 6548,\lambda
6584$, [SII]$\lambda 6716,\lambda 6731$ known as forbidden lines,
are expected. Taking the temperature $10^{4}$ $K$  which is
convenient for optical radiation, the density of the medium, the
energy of the supernova explosion, interstellar absorption,
neutral Hydrogen column density, shock wave velocities and  all
other related physical parameters are derived from these lines.

\section[]{Results}

In Figure 3, we present [SII]-[SII]cont./H$\alpha$-H$\alpha$ cont.
image of the remnant G85.9-0.6. Continuum substraction and image
division are done with IRAF {\it imarith} package. {\it Imarith}
package subtracts and divides the number of photons in each pixel
of the image obtained with the above mentioned filters. Calculated
[SII]/H$\alpha$ ratios and their errors in parenthesis taken from
10 different regions on the image are listed in Table 3, the
average of these ratios is 0.42$\pm$0.04.

Figure 4 presents the correlation between H$\alpha$ emission and
the radio emission of G85.9-0.6 at 1.4 GHz (radio contours are
obtained from http://skyview.gsfc.nasa.gov/cgi-bin/query.pl). The
western diffuse emission is well correlated with the radio
emission.

The flux values of the spectra shown in Figure 2 are obtained by
normalizing them with respect to H$\alpha$ flux value. With these
flux values and taking the electron temperature at $\sim 10^4$ K,
some basic parameters of SNR G85.9-0.6 are calculated and listed
in Table 4. The fluxes corrected for interstellar extinction are
normalized to I(H$\alpha) = 100$. For three region,
[SII]/H$\alpha$, [SII]$\lambda\lambda 6716 /6731$, [OIII]/H$\beta$
ratios and the parameters that are obtained from these ratios
found to be comparable. The average values are considered while
discussing the results.

To find these parameters first, through [SII]$\lambda\lambda 6716
/6731$ flux ratio the electron density is calculated
\citep{ost89}. Shock velocity is estimated from the ratio of
[OIII]/H$\beta$ as described by \citep{cox85, ray88, har87}.

Assuming that the remnant is still in the adiabatic phase of its
evolution, the pre-shock cloud density $n_{c}$ can be calculated
by using the relationship $n[SII]=45 n_{c}~(Vs/100)^2~cm^{-3}$
\citep{dop79, fes80}. Where n[SII] is the density calculated using
the [SII] emission line from the optical spectra. This equation is
valid if the transverse magnetic field is sufficiently low such
that the magnetic pressure does not dominate the post-shock
pressure as is discussed by \citet{dop96}.

Shock wave energy E is given by the equation $2 \times
10^{46}~\beta^{-1}~n_{c}~(Vs/100)^2 r_{s}^3~ergs$ \citep[see
e.g.][]{mck75} where $r_{s}$ is the radius of the shock wave. The
factor $\beta$ is approximately equal to 1 at the blast wave
shock.

The interstellar extinction $(E(B-V))$, absorption $(Av)$, and
neutral hydrogen column density $(N(HI))$ are calculated through
the relations $E(B-V)=0.664 c$, $Av=3.1 \times E(B-V)$ and
$N(HI)=5.4 \times 10^{21}\times E(B-V)$ \citep{kal76},
\citep{all84} and \citep{pre95}. Where $c$ is ${1} / {0.331}~log
[(H\alpha /H \beta )/3]$.

Furthermore, [SII]/H$\alpha$ ratio is also calculated from the
spectrum. In this calculation, $\lambda$6716 and $\lambda$6731
doublet line fluxes of [SII] is used. This ratio, for three
coordinates whose $\alpha$ and $\beta$ are given in the Table 2,
is calculated at its average value to 0.46.

\section{Discussion and Conclusions}
In this work, the imaging and spectroscopic observations of the
central part ($13 ^{'}\times 13 ^{'}$) of the galactic supernova
remnant G85.9-0.6 is observed for the first time in optical band.
The images show diffuse emission structure. As seen from the
Figures 1,3, the dominant emission extends from south-west to
north. The diffuse emission centered at $\alpha=20^{h} 59^{m}
00^{s}$, $\delta= 44^{0} 52^{'} 50^{''}$ that covers an area $6
^{'}\times8 ^{'}$ is very well correlated with radio emission (see
Fig.4).

As seen from the Tablo 3, [SII]/H$\alpha$ ratios take values
between 0.2-0.6. This ratio is $\sim$0.5, 0.6 on the places where
dominant diffuse emission comes indicating ionization of the shock
heated gas resulted from collision. On the other hand, the ratio
of $\sim$0.2-0.3 on the places where emission is faint indicating
that resulted emission originate from photoionization mechanism.
[SII]/H$\alpha$ ratio obtained from imaging is at 0.42 which is in
agrement with spectral measurements of 0.46. These values imply
that diffuse emission of G85.9-0.6 may be associated with HII
region.

From the spectra, we see the presence of H$\alpha$, H$\beta$,
[OIII]$\lambda 4959,\lambda 5007$, [NII]$\lambda 6548,\lambda
6584$ and [SII]$\lambda 6716,\lambda 6731$, in addition a weaker
[OI]$\lambda 6300$ and [OIII]$\lambda 4363$ emission lines. The
presence of [OI]$\lambda 6300$ line emission indicates that the
emission is originating from shock heated gas. Theoretical models
predict electron temperature through [OIII]J($\lambda
4959+5007/\lambda 4363$)ratio. Since [OIII]$\lambda 4363$ emission
line is very close to Hg I $\lambda 4358$, it is almost impossible
to differentiate them \citep{ost89}. Therefore, the electron
temperature (T) assumed to be $10^{4}$ $K$ in this work.

We detected a weak [OIII]/H$\beta$ emission line ratio at 1.18
indicating an oxygen-deficient remnant. Theoretical models of
\citet{cox85} and \citet{har87} give that for shocks with complete
recombination zones this value is $\sim$6, while this limit is
exceeded in case of shocks with incomplete recombination zones
\citep{ray88}. Our values measured from spectra indicates that
complete recombination takes place at those regions. Therefore,
considering our measurements and the theoretical models mentioned
above we estimated the velocity of the shock to be $80$ $km
s^{-1}$.

Since the G85.9-0.6 lies in the region between the local and
Perseus spiral arms, it has a reliable distance, of 5 kpc
\citep{kot01}. For this distance and $\theta=24^{'}$ we calculated
shock radius ($r_s$) to be $\sim$ $17.4$ $pc$. Explosion energy
and hydrogen column density of the G85.9-0.9 for this radius and
estimated shock velocity is calculated, to be $\sim9.2\times
10^{50}$ $ergs$ and $\sim1.53\times 10^{21}$ $cm^{-3}$
respectively. This much energy can be considered as the typical
energy released from a supernova explosion.

Typical value of the sulfur line ratio [SII]$(\lambda 6716 /
\lambda 6731)$ for SNRs should be close to 1.5 \citep{fra02}. This
ratio is calculated to be 1.13 indicating low electron density
which is calculated as 395 $cm^{-3}$ for this region.

We note here that, providing rather long exposure times and
examining its surrounding region will give more information about
the morphology and angular size of the G85.9-0.6 in optical band.
The images taken with different interference filters like [NII],
[OIII] will give more information about the inhomogeneities and
density variations of the medium.

\section*{Acknowledgments}

We thank TUBITAK for its support in using RTT150 (Russian-Turkish
1.5-m telescope in Antalya) under project number 09ARTT150-454-0.
A.S. is supported by TUBITAK Post-Doctoral Fellowship. This work
is supported by the Akdeniz University Scientific Research Project
Management and by T\"{U}B\.{I}TAK under project code 106T310. The
authors also acknowledge the support by Bogazici University
Research Foundation via code number 06HB301.

\clearpage

\begin{figure}[h]
\begin{center}
 \includegraphics[width=14cm]{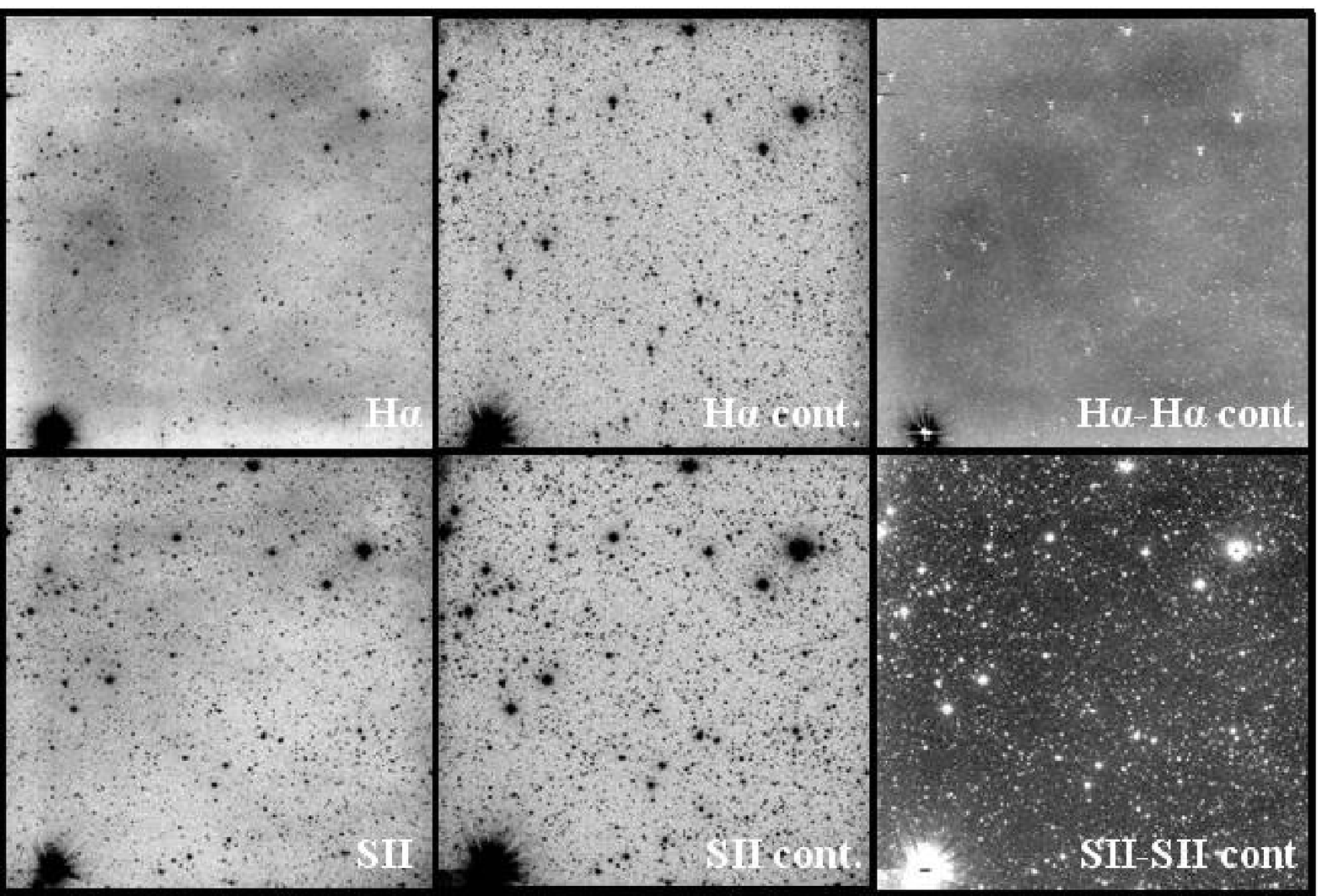}\\
\caption{H$\alpha$, H$\alpha$ cont., H$\alpha$-H$\alpha$ cont.,
[SII], [SII] cont. and [SII]-[SII] cont. images of G85.9-0.6. Each
field of view is 13.5 arcmin x 13.5 arcmin}
\end{center}
\end{figure}

\clearpage

\begin{figure}[h]
\begin{center}
\includegraphics[width=18cm]{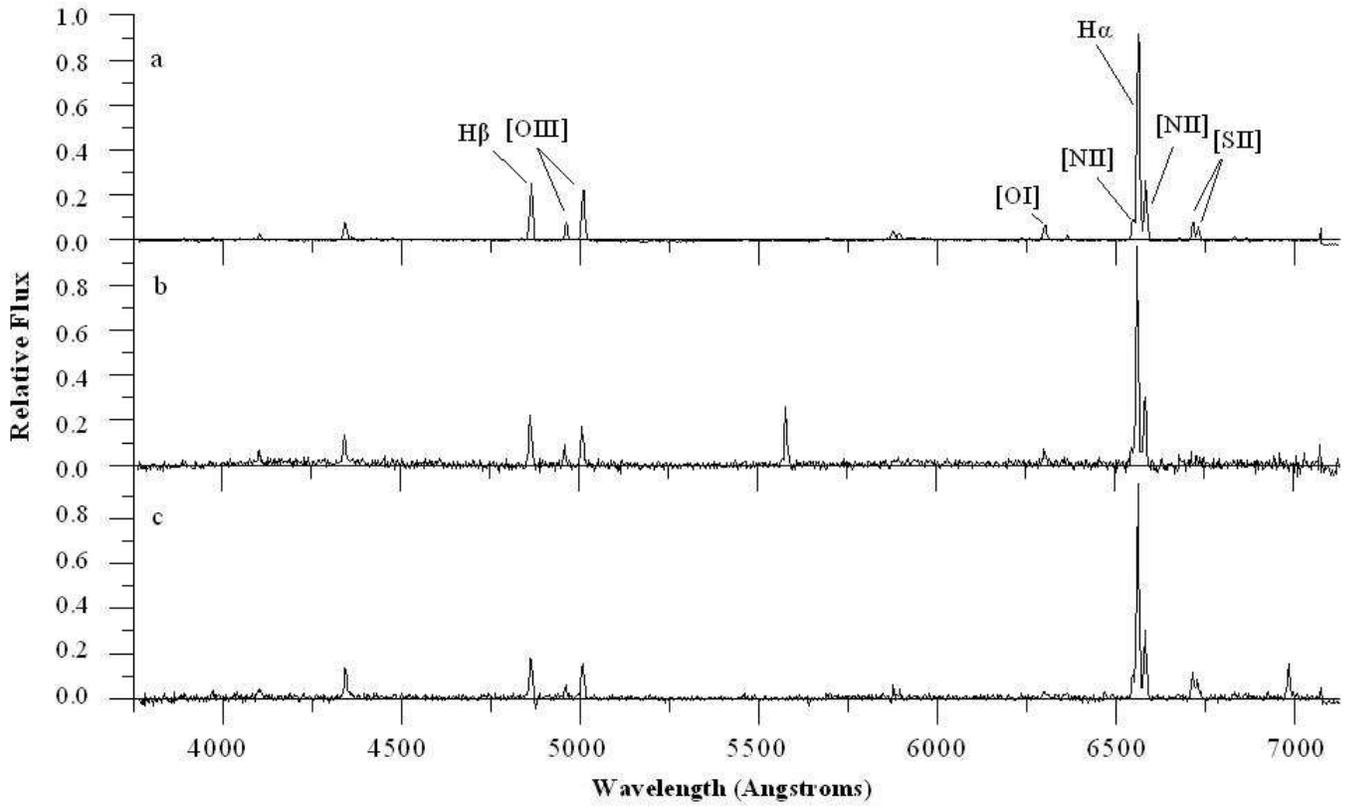}\\
\caption{The long slit spectra of G85.9-0.6 for area 1 (a), area 2
(b) and area 3 (c)} \label{fig:poodles}
\end{center}
\end{figure}

\clearpage

\begin{figure}[h]
\begin{center}
\includegraphics[width=18cm]{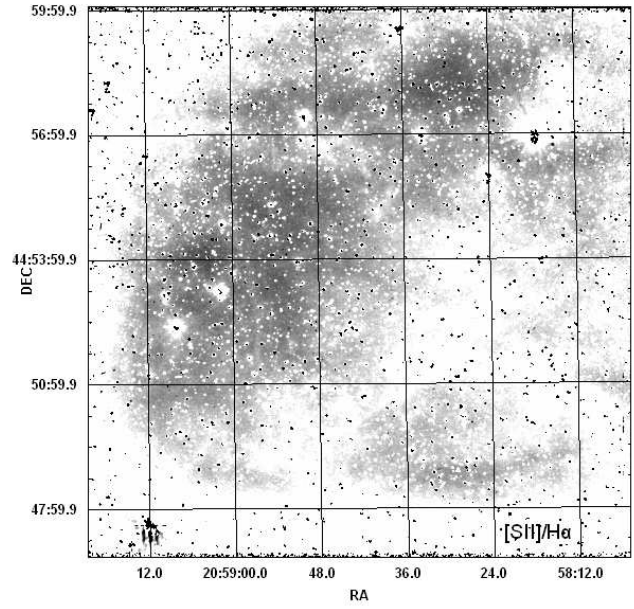}\\
\caption{The negative of the [SII]/H$\alpha$ image of G85.9-0.6.
The image has been smoothed to suppress the residuals from the
imperfect continuum subtraction. Field of view is 13.5 arcmin x
13.5 arcmin} \label{fig:poodles}
\end{center}
\end{figure}

\clearpage

\begin{figure}[h]
\begin{center}
\includegraphics[width=14cm]{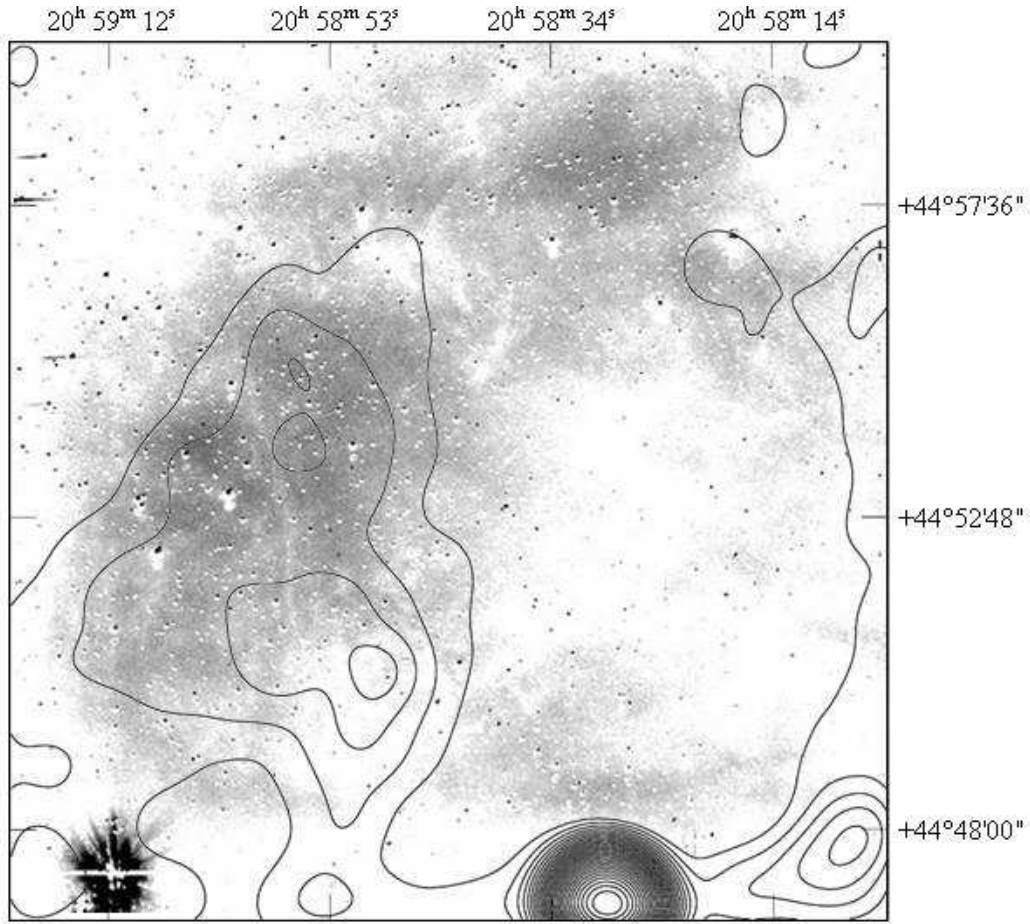}\\
   \caption{NVSS 1.4-GHz VLA radio continuum map superposed on the continuum
subtracted and smoothed H$\alpha$ image of G85.9-0.6}
 \label{fig:poodles}
\end{center}
\end{figure}

\clearpage

\begin{table}[h]
\begin{center}
\caption{The journal of observations. \label{tbl-1}}
\begin{tabular}{@{}cccc}
\tableline
Filter & \multicolumn{1}{c} {Wavelength} & \multicolumn{1}{c}{Obs.
Date (UT)}& \multicolumn{1}{c}{Exp. times (s)}\\
&(FWHM) ($\AA$) &  & \\ \hline
\hline
H$\alpha$       & 6563 (88)    & 01 June 2009   & 900 \\
H$\alpha$ cont. & 6446 (130)   & 01 June 2009   & 900 \\
SII             & 6728 (70)    & 01 June 2009   & 900 \\
SII cont.       & 6964 (300)   & 01 June 2009   & 900 \\
\hline
\end{tabular}
\end{center}
\end{table}

\clearpage

\begin{table}[h]
\begin{center}
\caption {The spectroscopic observations. \label{tbl-2}}
\begin{tabular}{@{}ccc@{}} \hline
Area & Slit Center (J2000) & Exposure times (s) \\
\hline
\hline
1 & $\alpha=20^{h} 59^{m} 04^{s}$ &  600\\
 & $\delta=44^{0} 52^{'} 19^{''}$ &  \\
 2 & $\alpha=20^{h} 58^{m} 49^{s}$ &  600\\
& $\delta=44^{0} 55^{'} 32^{''}$ &  \\
 3 &$\alpha=20^{h} 58^{m} 32^{s}$ &  600\\
& $\delta=44^{0} 58^{'} 24^{''}$ &  \\
\hline
\end{tabular}
\end{center}
\end{table}

\clearpage

\begin{table}[h]
\begin{center}
\caption {[SII]/H$\alpha$ ratios and their errors for different
regions of the G85.9-0.6. \label{tbl-3}}
\begin{tabular}{@{}ccc}
\hline
  Area & coordinate J(2000) & [SII]/H$\alpha$\\
   & $\alpha$ , $\delta$ & \\ \hline
    \hline
1&$20^{h} 58^{m} 29^{s} $ ~,~ $44^{0} 58^{'} 37^{''} $& 0.6 ($\pm$ 0.1)\\
2&$20^{h} 58^{m} 33^{s}$ ~, ~$44^{0} 58^{'} 30^{''}$& 0.5 ($\pm$ 0.1) \\
3&$20^{h} 58^{m} 49^{s} $ ~,~$44^{0} 55^{'} 07^{''} $ &0.5 ($\pm$ 0.1)\\
4&$20^{h} 58^{m} 53^{s}$ ~,~ $44^{0} 53^{'} 27^{''} $&0.5 ($\pm$ 0.1)\\
5&$20^{h} 58^{m} 48^{s}$ ~, ~$44^{0} 55^{'} 20^{''} $&0.4 ($\pm$ 0.1)\\
6&$20^{h} 58^{m} 29^{s}$ ~,~ $44^{0} 48^{'} 20^{''} $&0.3 ($\pm$ 0.1)\\
7&$20^{h} 58^{m} 18^{s}$ ~,~ $44^{0} 56^{'} 02^{''} $&0.3 ($\pm$ 0.1)\\
8&$20^{h} 58^{m} 11^{s}$ ~, ~$44^{0} 51^{'} 17^{''} $&0.2 ($\pm$ 0.1)\\
9&$20^{h} 59^{m} 03^{s}$ ~,~ $44^{0} 57^{'} 29^{''} $&0.3 ($\pm$ 0.1)\\
10&$20^{h} 59^{m} 04^{s}$ ~, ~$44^{0} 53^{'} 53^{''} $&0.6 ($\pm$ 0.1)\\
\hline
\end{tabular}
\end{center}
\end{table}

\clearpage
\begin{landscape}
\begin{table}[h]
\begin{center}
\caption {This table gives the relative line fluxes and the
parameters obtained from spectra for three different areas of the
G85.9-0.6. (F) shows fluxes uncorrected for interstellar
extinction,  (I) shows fluxes corrected for interstellar
extinction. Numbers in parentheses represent the signal-to-noise
ratio of the quoted values. All fluxes normalized F($H_{\alpha}$)=
100 and I($H_{\alpha}$)=100. The errors of the emission line
ratios and other parameters are calculated through standard error
calculation.\label{tbl-4}}
\footnotesize{
\begin{tabular}[10pt]{@{}lcccccc|lccc@{}}
 \hline
  Lines ($\AA$)&&& \multicolumn{3}{c} {Flux (F(H$\alpha$)=100, I(H$\alpha$)=100)}&&Parameters
  &\multicolumn{3}{c} {Values}\\
      & \multicolumn{2}{c} {Area 1}&\multicolumn{2}{c}{Area 2}&\multicolumn{2}{c}{Area 3} & & & & \\
& F&I(S/N)&F &I(S/N) & F & I(S/N) & &  Area 1&Area 2 &{Area3} \\

\hline \hline

4861 H$\beta$  & 28 & 52(58) & 25 & 48(14)  & 20 & 34(26) &[SII]/H$\alpha$     & 0.44$\pm 0.01$ & 0.45$\pm 0.01$  & 0.50$\pm 0.09$ \\
4959 [OIII]    & 8  & 15(40) & 11 & 21(7)   &8   & 13(9)  &[SII]$\lambda$6716/6731 & 1.09$\pm 0.08$ & 1.14$\pm 0.03$  & 1.17$\pm 0.2$ \\
5007 [OIII]    & 24 & 44(54) & 19 & 35(10)  &18  & 29(23) &[OIII]/H$\beta$         & 1.14$\pm 0.14$ & 1.16$\pm 0.20$  & 1.23$\pm 0.16$ \\
6300 [OI]      & 7  & 17(28) & 9  &13(4)    &5   &6(4)    &Ne (cm$^{-3}$)          & 470$\pm 120$   & 380$\pm 40$     & 336$\pm 26   $   \\
6548[NII]      & 10 & 10(28) & 10 &11(5)    &13  &14(11)  &n$_{c}$ (cm$^{-3}$)     &16$\pm 3$       & 13$\pm 1$        & 12$\pm 1$   \\
6563 H$\alpha$ & 100&100(245)&100 &100 (30) &100 &100(124)&c                       & 0.23$\pm 0.02$ &   0.38$\pm 0.04$ & 0.67$\pm 0.07$  \\
6584[NII]      & 29 & 33(56) &33  &31(9)    &33  &32(31)  &E(B-V)                  &0.15$\pm 0.02$  & 0.25$\pm 0.03$  & 0.45$\pm 0.05$ \\
6716 [SII]     & 25 &23(40)  & 24 &24(8)    &23  &27(7)   &Av                      & 0.47$\pm 0.05$ & 1.18$\pm 0.12$  & 1.40$\pm 0.14$  \\
6731 [SII]     & 20 &21(35)  & 20 &21(7)    & 19 &23(4)   &                        &                &                 & \\
\hline
\end{tabular}}
\end{center}
\end{table}
\end{landscape}

\end{document}